\documentclass{aa}
\usepackage{graphicx}
\def\la{\mathrel{\hbox{\rlap{\hbox{\lower4pt\hbox{$\sim$}}}\hbox{$<$}}}}
\def\ga{\mathrel{\hbox{\rlap{\hbox{\lower4pt\hbox{$\sim$}}}\hbox{$>$}}}}
\begin{document}

\title{Numerical simulations of expanding supershells in dwarf irregular
galaxies \\
II. Formation of giant HI rings}
\author{E. I. Vorobyov\inst{1,2} and Shantanu Basu\inst{2}\\ }

\offprints{E. I. Vorobyov}
\institute{ Institute of Physics, Stachki 194, Rostov-on-Don, Russia\\
\email{eduard\_vorobev@mail.ru}
\and
Department of Physics and Astronomy, University of Western Ontario, London,
Ontario,  N6A 3K7, Canada
\email{basu@astro.uwo.ca}
}
\date{}

\abstract{We perform numerical hydrodynamic modeling of various physical
processes that can form an HI ring as is observed in Holmberg~I (Ho~I).
Three energetic mechanisms are considered: multiple
supernova explosions (SNe), a hypernova explosion associated with a gamma
ray burst (GRB), and the vertical impact of a high
velocity cloud (HVC). The total released energy has an upper limit of $\sim
10^{54}$~ergs.  We find that multiple SNe are in general more effective
in producing shells that break out of the disk 
than a hypernova explosion of the same total energy. 
As a consequence, multiple SNe form rings with a high ring-to-center
contrast ${\cal K}\la 100$ in the HI column density, whereas 
single hypernova explosions form rings with ${\cal K}\la 10$.
Only multiple SNe can reproduce both
the size (diameter $\sim 1.7$~kpc) and the ring-to-center contrast (${\cal K}\sim 15-20$) 
of the HI ring in Ho~I. High velocity clouds create HI rings 
that are much smaller in size ($\la 0.8$~kpc) and contrast (${\cal K}\la
4.5$) than seen in Ho~I. We construct model position-velocity (pV) diagrams and find that they
can be used to distinguish among different HI ring formation mechanisms.
The observed pV-diagrams of Ho~I (Ott et al.~\cite{Ott}) are  best
reproduced by multiple SNe. We conclude that the giant HI ring in
Ho~I is most probably formed by multiple SNe. We also find that the appearance
of the SNe-driven shell in the integrated HI image depends on
the inclination angle of the galaxy. In nearly face-on galaxies, the integrated
HI image shows a ring of roughly constant HI column density 
surrounding a deep central depression, whereas
in considerably inclined galaxies ($i>45^\circ$) the HI image is characterized
by two kidney-shaped density enhancements and a mild central depression.
}

\titlerunning{Formation of giant HI rings}
\authorrunning{E. I. Vorobyov, S. Basu}
\maketitle

\section{Introduction}
There are a few nearby low-mass dwarf irregular galaxies  (dIrr's) such as
Holmberg~I (Tully et al. \cite{Tully},
Ott et al. \cite{Ott}), M81~dwA (Sargent et al. \cite{Sargent}, Westpfahl \& Puche \cite{Westpfahl}), Sagittarius~DIG (Young \& Lo \cite{Young}), Sextans~A (Skillman et al.
\cite{Skillman}, Stewart \cite{Stewart}), the HI morphology of
which is totally dominated by a single ring structure of size comparable
to or bigger than their optical extent. 
The integrated HI image of these dIrr's shows a central HI hole surrounded
by a denser HI ring. 
The contrast in the azimuthally averaged HI column density between  the central hole
and the ring varies from $\sim 3-4$ in Sextans~A and M81~dwA to $\sim15-20$ in Holmberg~I.
HI rings that do not dominate the overall HI extent of a
galaxy, yet occupy a considerable region (kpc scale),
have also been found in other dIrr's such as DDO~47 (shell~13, according to Walter
\& Brinks \cite{Walter}) and Holmberg~II (shell~21, according to 
Puche et al. \cite{Puche}).

There are a number of scenarios that could create HI holes in galactic disks 
(see S\'anchez-Salcedo
\cite{SS}), which can loosely be divided into two groups comprising 
the energetic and non-energetic mechanisms.
The energetic mechanisms assume a deposition of vast amount of energy into
the interstellar medium and  include multiple supernova explosions 
(Mac Low \& McCray \cite{MacLow2}, De Young \& Heckman \cite{deYoung},
Mac Low \& Ferrara \cite{MacLow}, Silich et al. \cite{Silich}),  
impact of high velocity clouds (Tenorio-Tagle et al. \cite{TT}, Comer\'on
\& Torra \cite{Comeron}, Rand \& Stone \cite{Rand}),
and gamma ray bursts (Efremov et al. \cite{Efremov}, Loeb \& Perna \cite{LP}).
The non-energetic mechanisms include a combined action of thermal and gravitational 
instabilities in the gas disk (Wada et al. \cite{Wada2}), turbulent clearing 
(Elmegreen \cite{Elmegreen}, Walter \& Brinks \cite{Walter2}), and
ultra-violet erosion of the HI disk (Vorobyov \& Shchekinov \cite{VS}).
It appears however very unlikely that a central HI depression surrounded by 
a denser HI ring with size comparable to the galaxy's optical extent 
could be produced by a non-energetic mechanism. 
Indeed, Vorobyov et al. (\cite{Vor}) have recently shown that the HI ring-like
morphology of Holmberg~I (Ho~I) can be produced by multiple supernova explosions
(SNe).

In this paper, we consider two other energetic mechanisms of HI ring formation, 
namely the  impact of high velocity clouds (HVCs) and gamma ray bursts (GRBs),
and show that they cannot explain the formation of the observed giant HI ring 
in Ho~I. We generate the position-velocity diagrams and show that they can be used
to distinguish between the rings created by different energetic mechanisms.
We find that the appearance of HI rings created by multiple SNe 
is sensitive to the inclination angle of the galaxy. In considerably inclined
galaxies, the HI ring would rather appear as two kidney-shaped density
enhancements similar to those observed in Sectans~A.

The paper is organized as follows. In Sect.~\ref{model} the numerical hydrodynamic
model for simulating multiple SNe, GRBs, and the impact
of HVCs is formulated. A comparative study of multiple SNe
and GRBs of the same total energy is performed in Sect.~\ref{SNGRB}. The
collision of HVCs with the galactic gas disk is considered in Sect.~\ref{cloud}.
The main results are summarized in Sect.~\ref{sum}.

\section{Numerical model}
\label{model}
Our model galaxy consists of a rotating gas disk, stellar disk, and a 
spherically symmetric dark matter halo.  
The density profile of the stellar component is chosen as:
\begin{equation}
\rho=\rho_{\rm s0} \: {\rm sech}^2(z/z_{\rm s}) \: \exp(-r/r_{\rm s}),
\label{stellar}
\end{equation}
where $\rho_{\rm s0}$ is the stellar density in the center of the galaxy, and $z_{\rm s}$ and 
$r_{\rm s}$ are the vertical scale height and radial scale length of the stellar component, respectively. 
We assume that the density profile of the dark matter (DM) halo can be approximated 
by a modified isothermal sphere (Binney \& Tremaine \cite{Binney})
\begin{equation}
\rho_{\rm h}={\rho_{\rm h0} \over 1+(r/r_{\rm h})^2},
\end{equation}
where the central density $\rho_{\rm h0}$ and the characteristic scale length $r_{\rm h}$ 
were given by Mac Low \& Ferrara (\cite{MacLow}) and Silich \& Tenorio-Tagle (\cite{Silich}) 
based on the study of the dark-to-visible mass ratios by Persic, Salucci, \& Stel (\cite{Persic}):
\begin{equation}
\rho_{\rm h0}=6.3 \times 10^{10} \left( {M_{\rm h} \over M_{\odot}}
\right)^{-1/3}~h^{-1/3}~M_{\odot}~{\rm kpc}^{-3},
\label{halo1}
\end{equation}
\begin{equation}
 r_{\rm h}=0.89 \times 10^{-5} \left( {M_{\rm h} \over
 M_{\odot}}\right)^{1/2}~h^{1/2}~{\rm kpc}.
 \label{halo2}
\end{equation}
Here, $h$ is the Hubble constant in units of 100~km~s$^{-1}$~Mpc$^{-1}$ and 
$M_{\rm h}$ is the {\it total} halo mass. We adopt $h=0.65$ throughout the paper.

Our model galaxy is meant to represent Ho~I and we use the observations
of Ott et al. (\cite{Ott}) to constrain the parameters of our model.
We note that $M_{\rm h}$ in Eqs.~(\ref{halo1}) and (\ref{halo2}) is, in
fact, the total mass of the halo. We vary $M_{\rm h}$ until the actual halo
mass confined within the HI diameter of Ho~I 
(5.8~kpc, Ott et al. \cite{Ott}) agrees with the observed value, $3.1\times
10^8~M_\odot$.  
This fit yields a value of $M_{\rm h}=6.0\times 10^9~M_\odot$ for Ho~I.
Once the total halo mass is fixed, we derive the parameters of the halo
density distribution using Eqs.~(\ref{halo1}) and (\ref{halo2}), which
are further used to compute the gravitational potential of the halo
as described in Vorobyov et al. (\cite{Vor}). 
We have adopted a value of $z_{\rm s}=300$~pc, which is typical
for dwarf irregular galaxies.
The radial scale length
$r_{\rm s}=1.7$~kpc of the stellar disk is estimated from the $I_{\rm
c}$-band radial surface brightness profile of Ho~I. 
With $r_{\rm s}$ and $z_{\rm s}$ being 
fixed, the stellar density $\rho_{\rm s}^0$ in Eq.(~\ref{stellar}) is varied so as to 
obtain the measured luminous stellar mass of $M_{\rm s}=1.0 \times 
10^8~M_{\odot}$. This results in $\rho_{s}^{0}\approx0.02~M_{\odot}$~pc$^{-3}$. 
The adopted parameters of the stellar disk are further used
to compute the stellar gravitational potential by solving the Poisson equation.

\begin{figure}
    \resizebox{\hsize}{!}{\includegraphics{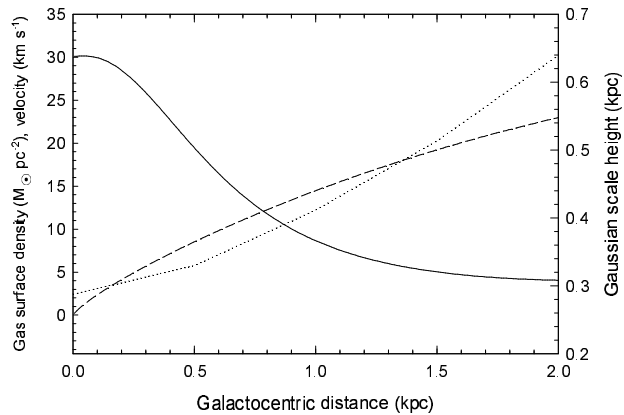}}
      \caption{The equilibrium radial distribution of gas surface density
      (the solid line), the initial rotation curve (the dashed line), and
      the Gaussian vertical scale height of gas distribution (the dotted
      line). }
         \label{fig0}
\end{figure}

Once the stellar and DM halo gravitational potentials are fixed, 
we obtain the initial gas density distribution by solving the steady-state momentum equation 
as described in Vorobyov et al. (\cite{Vor}). We set the gas velocity dispersion
$\sigma=(R T/\mu)^{1/2}$ to be 9~km~s$^{-1}$ (Ott et al. \cite{Ott}).
We vary the rotation curve until
the initial gas surface density distribution becomes exponential, which
is in agreement with observations of the HI radial distribution in many
dIrr's (Taylor et al. \cite{Taylor}). 
The observed rotation curve (RC) of Ho~I is known to the accuracy
of the inclination angle, which in turn is poorly determined (Ott et al.~\cite{Ott}) 
due to Ho~I's small inclination. We have chosen the initial 
RC in our model so that the initial radial gas surface 
density distribution well reproduces the observed profile at $r>1$~kpc, 
i.e. at radii which are not affected by the 
subsequent ring expansion. We do not expect our initial RC to 
match the currently observed RC, since the latter
is already affected by ring formation. Furthermore,
accurately modeling the observed RC 
(for an assumed inclination angle) requires non-axisymmetric simulations, 
since the dynamical center of Ho~I is $\sim 0.7$~kpc
offset from the morphological center of the ring. 
This offset may complicate the shape of the observed RC, 
because it is measured around the dynamical center. This task is beyond
the scope of our paper.
The resulting radial gas surface density
profile, as well as the Gaussian vertical scale height $h$ of the gas distribution
and initial gas rotation curve, are plotted in Fig.~\ref{fig0} by the solid,
dotted, and dashed lines, respectively. 
The total gas mass
within the computational domain is $M_{\rm gas}=1.0\times 10^8~M_\odot$,
of which $\sim 30\%$ is contributed by He ($M_{\rm HI+He} \approx 1.4\, M_{\rm
HI}$, $\mu=1.27$, Brinks \cite{Brinks}). This value of $M_{\rm gas}$ roughly agrees with that
obtained by Ott et al. (\cite{Ott}).

A usual set of hydrodynamical equations in cylindrical coordinates (with the assumption of axial symmetry) 
is solved using the method of finite-differences with a time-explicit, operator-split solution procedure as
used in the ZEUS-2D code described in detail in Stone \& Norman (\cite{SN}). 
The computational domain spans the range of $7.2~{\rm kpc}\times 2~{\rm
kpc}$ in the vertical and horizontal directions, respectively, with a resolution of 5~pc.
We have implemented the optically thin cooling curve given 
in Wada \& Norman (\cite{Wada}) for a metallicity of one tenth of solar, which is typical for dIrr's. 
The cooling processes taken into account are: 
(1) recombination of H, He, C, O, N, Si, and Fe; 
(2) collisional excitation of HI, CI-IV, and OI-IV; 
(3) hydrogen and helium bremsstrahlung; 
(4) vibrational and rotational excitation of $\rm H_2$; 
(5) atomic and molecular cooling due to fine-structure emission of C, C+, 
and O, and rotational line emission of CO and $\rm H_2$.
We use an empirical heating function tuned to balance the cooling in 
the background atmosphere so that it maintains the gas in hydrostatic 
and thermal equilibrium; it may be thought of as a crude model for 
the stellar energy input. However, heating is prohibited at $T_{\rm gas}>2.0\times10^4$~K 
to avoid the effects of spurious heating of a bubble's interior
by the time-independent heating function. This 
is physically justified since most of the heating in the warm interstellar medium 
comes from the photoelectric heating of polycyclic aromatic hydrocarbon molecules 
(PAHs) and small grains, which will be either evaporated or highly ionized in the bubble filled with 
hot supernova ejecta. 
Cooling and heating are treated numerically at the end of the time integration
step using an implicit update to the energy equation. The implicit equation
for energy density is solved by Newton-Raphson iteration, supplemented by a bisection algorithm for 
occasional zones where the Newton-Raphson method does not converge. In order to monitor accuracy, 
the total change in the internal energy density in one time step is kept below 
$15\%$. If this condition is not met, the time step is reduced and a solution 
is again sought. In the following we give a brief description on the mechanisms of energy
injection used in our numerical simulation. 

\section{Energetic mechanisms of HI ring formation}
Here, we briefly review possible mechanisms of HI ring
formation as observed in Ho~I. \\
{\it Consecutive supernova explosions}. The origin of the giant HI rings in dIrr's is traditionally thought to lie in the
combined effect of stellar winds and supernova explosions produced by young
stellar associations (see e.g. Ott et al. \cite{Ott}). 
In our simulations, the energy of supernova explosions is released in the form of thermal energy 
in the central region with a radius of 30~pc filled with the hot ($T\sim10^7~K$) 
and rarified ($n\sim 10^{-3}$~cm$^{-3}$) gas, which is presumably formed
by the previous action of stellar winds.
We use  a constant wind approximation, i.e. at each time step we add energy
to the source region at a rate of $\dot{E}=L$, where $L$ is the mechanical
luminosity defined as the total released energy of SNe divided by the duration of the energy
input phase. We choose the energy input phase to last for 30~Myr, 
which corresponds roughly to the difference in the lifetimes of the most and 
least massive stars capable of producing SNe in a cluster of simultaneously 
born stars. Since in the present simulations we deal with large stellar 
clusters with hundreds of supernovae, the release of the energy of SNe
in the form of thermal energy is justified (Mac Low \& McCray \cite{MacLow2}).

{\it Gamma ray bursts}. The consecutive SN explosions may not be the only mechanism
that could release $10^{53}-10^{54}$~ergs of energy, 
enough to form giant HI rings.
Another mechanism has been suggested by Efremov et al. (\cite{Efremov})
and Loeb \& Perna (\cite{LP}), who argued that the GRB
explosions are powerful enough to make kpc-size shells in the interstellar
media of spiral and irregular galaxies. Although the physics of 
GRBs is still poorly understood, the general picture emerging is that they are
highly energetic events ($\le 10^{54}$~ergs, see e.g. Paczy\'{n}ski \cite{Pac}) 
that release energy in a short period of time (of the order of a few seconds). 
Consequently, we model the GRB explosion (hereafter, ``hypernova'' according
to Paczy\'nski \cite{Pac}) by an instantaneous release of
thermal energy within a sphere of 30~pc, filled with hot ($T\sim10^7~K$) 
and rarified ($n\sim 10^{-3}$~cm$^{-3}$) gas. We have also
explored the injection of hypernova energy in the form of kinetic energy
and found that it does not noticeably influence the dynamics of the hypernova-driven shell. 

{\it Impact of high velocity clouds}. Another possible mechanism that could
create HI holes in the galactic disks was proposed
by Tenorio-Tagle et al. (\cite{TT}), who argued that the infall of high velocity clouds
could deposit $10^{52}$ to $10^{54}$ ergs per collision. 
The numerical simulations of Tenorio-Tagle et al. (\cite{TT}) have indeed
shown that HVCs are capable of forming the giant curved arcs and cavities
in the Galactic disk (see also Rand \& Stone (\cite{Rand})
for numerical simulations of HVC impact in  NGC~4631). Most
previous numerical simulations have been concerned with the infall of HVCs
in massive galaxies, because their collisional cross section is much larger
than that of dwarf irregulars. Moreover, the gas disks in dwarf irregulars
are in general thicker than those of massive spiral galaxies, which would
make it more difficult for an HVC to penetrate the disk.
In summary, the formation of a giant HI hole
surrounded by a denser HI ring (as is observed in Ho~I and other dwarf irregulars) 
by the infall of HVCs is not obvious and requires further investigation.

Taking into account the diameter of the HI ring in Ho~I ($\sim 1.7$~kpc), 
we consider the most energetic HVCs,
which cover a velocity range $200~{\rm km~s}^{-1}\le v_{\rm \scriptstyle
HVC}\le 300~{\rm km~s}^{-1}$
and have HI column density $10^{20}~{\rm cm}^{-3} \le N_{\rm HI} \le 10^{21}$~cm$^{-3}$. 
In our simulations the kinetic energies of HVCs range from $1.0\times 10^{53}$ to
$2.5 \times 10^{54}$~ergs, which corresponds to a variation in HVC masses of $2.5\times
10^5~M_\odot$ to $2.5 \times 10^6~M_\odot$.
Such massive and energetic HVCs most probably have an extragalactic origin.

\section{Supernova explosions versus gamma ray bursts}
\label{SNGRB}

\begin{figure}
   \resizebox{\hsize}{!}{\includegraphics{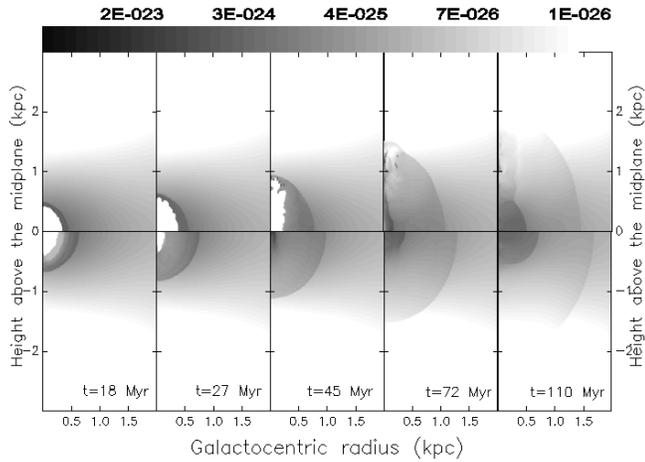}}
      \caption{Temporal evolution of the gas volume density  distribution
      after the release of $2\times 10^{53}$~ergs of thermal energy at $z=0,r=0$~pc. 
      The upper panels correspond to 200 consecutive SN explosions over
      30~Myr, while the lower panels show the impact of a single hypernova
      of the same total energy of $2.0\times 10^{53}$~ergs. The scale bar is in gm~cm$^{-3}$.}
         \label{fig1}
\end{figure}

\begin{figure}
    \resizebox{\hsize}{!}{\includegraphics{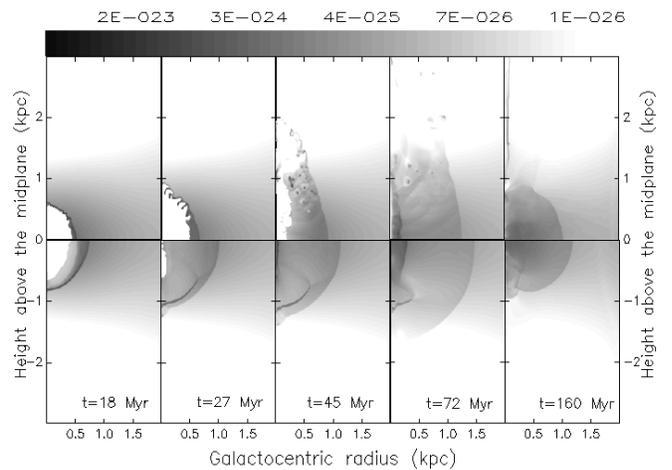}}
      \caption{The same as Fig.~\ref{fig1}, but for the energy input of
      $5\times 10^{53}$~ergs.}
         \label{fig2}
\end{figure}

\begin{figure}
    \resizebox{\hsize}{!}{\includegraphics{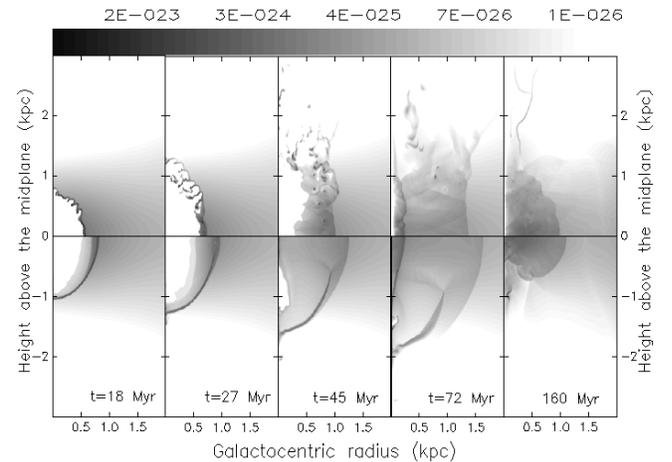}}
      \caption{The same as Fig.~\ref{fig1}, but for the energy input of
      $10^{54}$~ergs.}
         \label{fig3}
\end{figure}

We start by showing in Fig.~\ref{fig1} the temporal evolution of the distribution
of the gas volume density produced by 200 consecutive
SNe (the upper panels) and by a hypernova explosion of the same total
energy of $2\times 10^{53}$~ergs (the lower panels).  
The overall gas dynamics is similar
in both cases. However, there are minor differences seen not only in the dynamics
of the hot gas but also in the shape of the blown-up shell.
In the case of a hypernova explosion, there is no ``blowout'' observed
(i.e. the shell breaking out of the disk and pumping the hot gas into the
intergalactic medium).
The hot gas is always confined inside the shell in the early expansion
phase and it cools down at later times $t>40$~Myr due to the radiative cooling.
As a result, the shell collapses at $t=45$~Myr, 
creating a mildly compressed central core, the gas density of which 
is however below the Jeans limit. 
In the case of multiple SNe, part of the hot gas is lifted to a higher altitude
of $z \ge 1$~kpc due to a buoyancy effect developing at $t\approx 45$~Myr.
It is also seen that in the early expansion phase the shell is on average
thicker in the case of the hypernova explosion than in the case of multiple
SN explosions. Since the numerical modeling of Vorobyov et al. (\cite{Vor}) suggested
that the shell in Ho~I has already blown out of the disk, we do not consider 
further the energy release of $2.0\times 10^{53}$~ergs.

The difference in the shell dynamics between the multiple SN and hypernova
becomes pronounced as one considers
more energetic explosions. For instance, Fig.~\ref{fig2} shows the distribution
of the gas volume density produced by 500 consecutive
SN explosions (the upper panels) and by a single hypernova explosion of the
same total energy 
$5.0\times 10^{53}$~ergs (the lower panels). Now, the dynamics  of the SN-driven
shell shows a clear blowout phase at $t\approx 45$~Myr after the beginning
of the energy input phase. On the contrary, the hypernova-driven shell
never breaks out of the disk, though its energy is equal to the total energy
released by multiple supernovae. This difference remains for presumably
an upper limit energy release of $10^{54}$~ergs that a stellar cluster could produce in 
Holmberg~I. As seen in the
upper panels of Fig.~\ref{fig3}, the shell dynamics governed by 
1000 consecutive SN explosions exhibits a violent blowout, whereas
that governed by a hypernova of the same total energy of $10^{54}$~ergs shows
almost no sign of blowout. This is in good agreement with the previous
model of Efremov, Ehlerov{\'a} \& Palou{\v s} (\cite{EEP}), who
used a thin shell approximation to study the evolution of shells formed
by a single hypernova and multiple SNe. These authors
have also found that an abrupt energy input creates shells that do not
blow out to the galactic halo for energies $\la 10^{54}$~ergs.
Noticeably, the 
final fate of the gas distribution is similar in all cases: the hot bubble fills
in and the gas disk mostly recovers its pre-explosion appearance  after $\ge
160$~Myr with 
a slightly more centrally condensed radial gas distribution.

\begin{figure}
    \resizebox{\hsize}{!}{\includegraphics{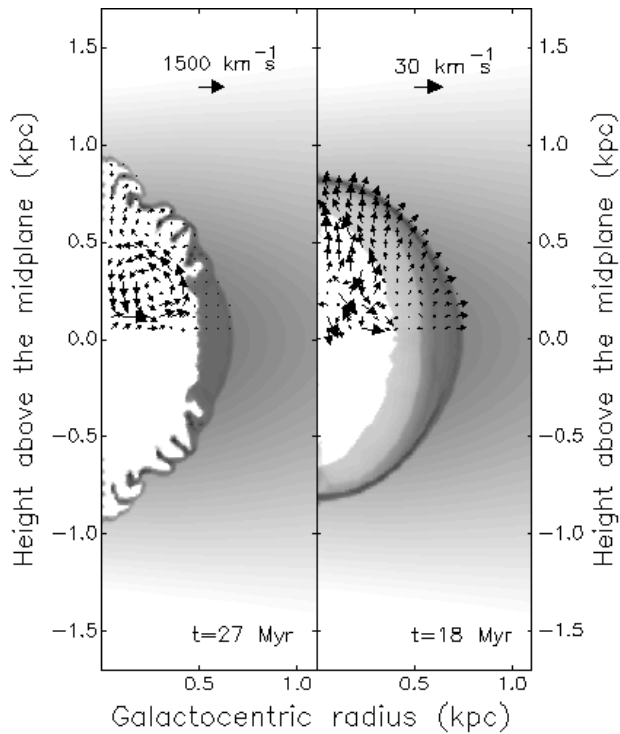}}
      \caption{The vertical cut through the distribution of the gas volume
      density produced by 500 consecutive SNe at t=27~Myr (the left panel) and single
      hypernova of the same total energy of $5.0\times 10^{53}$~ergs (the
      right panel). The gas velocity field is shown by the arrows.}
         \label{fig4}
\end{figure}

The occurrence of the blowout in the case of multiple supernova explosions
and its absence in the case of a hypernova explosion can
be understood if one considers the dynamics of the hot gas filling
the shell interior. 
In Fig.~\ref{fig4} we plot the gas volume density distribution and 
velocity field produced by 500 consecutive SN explosions
(the left panel) and a hypernova of the same total energy of $5.0 \times
10^{53}$~ergs (the right
panel). An obvious difference is seen: the hot gas ejected
by multiple SNe forms a ``vortex'' that acts to de-stabilize the swept-up shell 
of cold material via a Kelvin-Helmholtz instability. The shell loses its
smooth elliptical form and develops ``ripples'' at its tops. The dynamical 
pressure of the hot gas (note that its velocity is much higher 
than in the case of a hypernova explosion) accelerates
the ``rippled'' shell. As a consequence, the shell shows a strong Rayleigh-Taylor 
instability via the development of a characteristic `spike-and-bubble' morphology,
as is seen in the left panel of Fig.~\ref{fig4}.
In the case of a hypernova explosion of the same total energy, the gas velocity
field plotted in the right panel of Fig.~\ref{fig4} lacks any circular motion
inside the shell.
The smooth elliptical shell of cold material 
of roughly the same size as in the case of multiple SNe expands 
at roughly the sound speed.
The shell has already started losing its pressure support (as implied by the
thickening of the shell walls) due to radiative cooling of its hot interior.
As a consequence, the decelerating shell does not 
develop a Rayleigh-Taylor instability and the blowout phase is not observed.
We note here that a hypernova-driven shell expands out to a certain radius
faster than a shell created by multiple SNe with the same total energy, as reported by
Efremov et al. (\cite{EEP}).
Our numerical simulations have shown that the difference in the dynamics of 
shells produced by multiple SNe and a single
hypernova of the same total energy intensifies as one considers
more flattened gas systems. 
The maximum difference is found in the strongly vertically 
stratified gas distributions assumed for massive spiral galaxies.

The vortex forms since the shell becomes elongated in the vertical
direction. Consider the simplified situation of hot gas ejected radially
by SNe which is reflected from the cold dense walls of the elliptical shell.
The gas will not be reflected along the local normal direction to the shell
unless it is moving along $z=0$ or $r=0$. In general, a tangential
component of the velocity of hot gas $v_{\rm t}$ (with respect
to the shell's walls) is generated. It is always directed upwards
for gas above the midplane and downwards for gas below the midplane.
Let us consider the upper hemisphere.
The occurrence of $v_{\rm t} \ne 0$ near the shell walls and the
axial symmetry of the shell makes the hot gas 
(streaming upwards along the shell walls) accumulate
at the top of the shell, because it cannot pass through the symmetry axis.
Both the growing pressure of this hot gas (due to its negligible cooling) 
and the downward pull of the $z$-component of the galactic 
gravitational field lead to a downward flow
along the axis of symmetry.
This completes the circle and generates a vortex structure as seen in 
the left panel of Fig.~\ref{fig4}.
The development of such vortices is also seen in the axially symmetric numerical
simulations of Recchi et al. (\cite{Recchi}). However, a strong non-axisymmetry 
of the SNe-driven shell may complicate the formation of the vortex. 

The driving force for the vortex is a continuous or
quasi-continuous release of energy by SNe of a stellar cluster or a group
of closely located stellar clusters.
This is also the reason why vortices do not form in a shell created
by a single hypernova explosion. We have found that the vortex develops
even when the number of SNe is quite moderate ($\sim 20$) and the release
of energy is discrete.
The length scale of the vortex is approximately equal to the shell's
semi-minor axis and time scale is limited by the duration of energy
input from SNe, i.e. $\la 30 $~Myr.
In general, at least two vortices can co-exist within a single shell: one
in the upper hemisphere and the other in the lower hemisphere of the shell.

\begin{figure}
    \resizebox{\hsize}{!}{\includegraphics{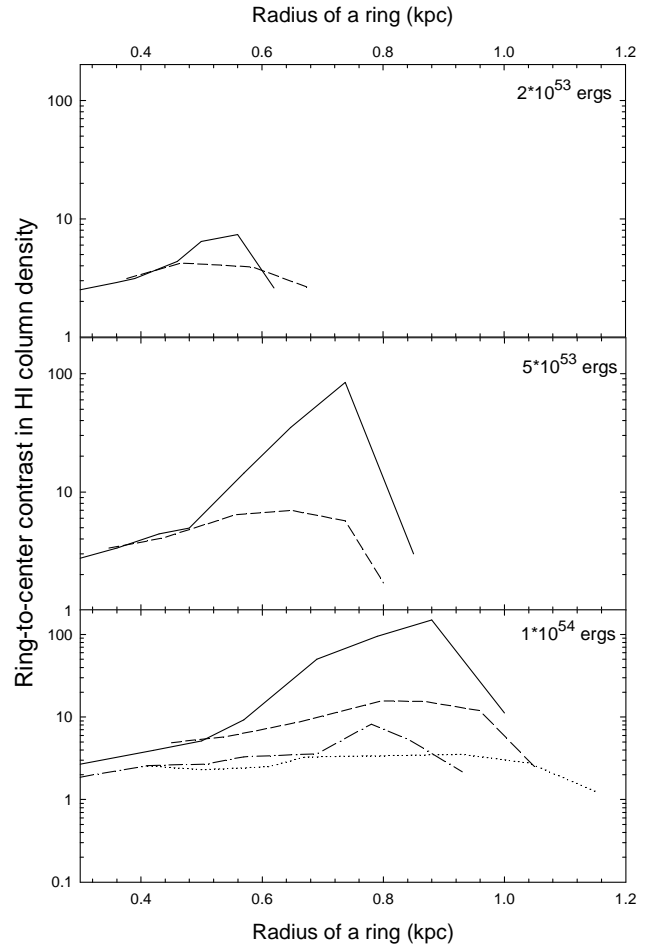}}
      \caption{The contrast in the HI column density between the ring and
      the central depression ${\cal K}=N_{\rm r}(HI)/N_{\rm c}(HI)$ 
      as a function of the ring radius. The solid 
      lines show the contrast $\cal K$ produced by multiple SNe, whereas
      the dashed lines show that produced by a single hypernova of the same
      total energy. An inclination angle of $5^\circ$ is assumed.
      The energy input
      is indicated in the right upper corner of each panel. The dashed-dotted
      and dotted lines in the lower panel give the contrast $\cal K$ for
      an inclination angle of $45^\circ$. }
         \label{fig5}
\end{figure}

Hypernovae tend to form complete shells. Such shells would appear in nearly face-on
dwarf galaxies as HI rings
with a low ring-to-center contrast in the HI column density.
For instance, in Fig.~\ref{fig5} we plot the contrast in the HI column density between 
the ring and the central depression, ${\cal K}=N_{\rm r}(HI)/N_{\rm c}(HI)$, as a
function of the ring radius $R$ obtained for three different energy inputs. 
The ring column density $N_{\rm r}(HI)$ is computed by azimuthally averaging
$N(HI)$ around the ring.
An inclination angle (i.e. the angle between the rotation axis of the
galaxy and the line of sight) of $i=5^{\circ}$ is assumed, which is appropriate
for a nearly face-on galaxy.
The solid lines give
the contrast ${\cal K}$ for the multiple SN explosions, whereas the dashed
lines do that for the hypernova explosion. As is seen, the ring-to-center
contrast ${\cal K}$ produced by the hypernova explosion never exceeds 10 even 
for the upper limit energy of
hypernova explosions of $10^{54}$~ergs. On the other hand, multiple SNe
can form HI rings with a much higher ring-to-center contrast ${\cal K} 
\la 100$. The latter takes place when the SN-driven shell is in the blowout phase. 
Note that the difference
in ${\cal K}$ between the HI rings produced by multiple SNe and those produced
by hypernovae  
smears out if the total released energy is $\le 2\times 10^{53}$~ergs.

\begin{figure}
    \resizebox{\hsize}{!}{\includegraphics{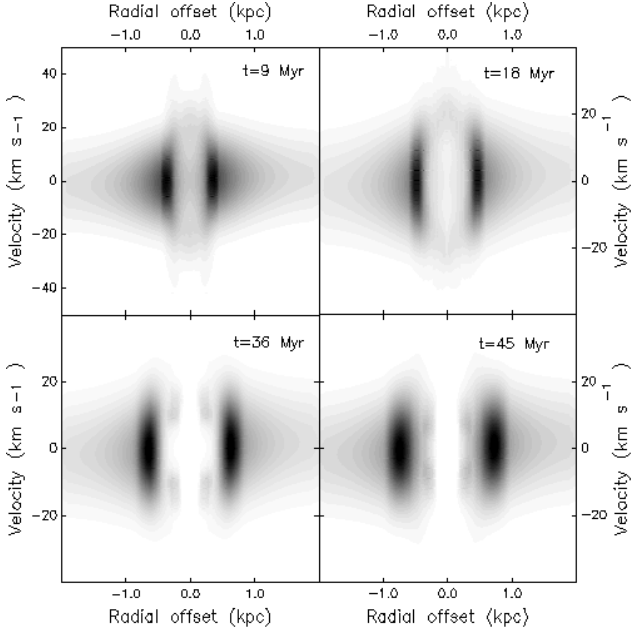}}
      \caption{The pV-diagrams of the shell created by 500 consecutive
      SNe. The pV-cuts are taken along the major axis of the model galaxy
      viewed at an angle of $5^{\circ}$. The evolutionary time of the shell
      is indicated in the right upper corner. The quantity $S_{\rm HI}$
      is plotted in all model pV-diagrams.}
         \label{fig6}
\end{figure}

\subsection{PV-diagrams}
\label{PVD}
The position-velocity (pV) diagram is a powerful tool to search for coherent structures
in the interstellar medium. 
In this section we compute model pV-diagrams in order to determine
their utility for distinguishing between the SN-driven shells and those produced by a 
hypernova explosion of the same total energy.  
The HI flux density $S_{\rm HI}$ is obtained using the following conversion
formula (Binney \& Merrifield \cite{BM}):
\begin{equation}
{M_{\rm HI} \over \triangle v} = 2.35 \times 10^5~\left( {D \over {\rm Mpc}} 
\right)^{2} \left({S_{\rm HI} \over {\rm Jy}}
\right) \ \   \left[{M_{\odot} \over {\rm km\: s^{-1}}}\right],
\label{HImass}
\end{equation}
where  for the distance $D$ we take 3.5~Mpc, appropriate for Holmberg~I.
The position-velocity cuts are taken along the major axis of the projected shell.
The width of the cuts is 30~pc. We tried  twice wider cuts (60~pc), but found little
difference in the appearance of the resulting pV-diagrams as compared to
those obtained with narrower cuts. The gas is assumed to be thermalized when
constructing the model pV-diagrams, i.e. a Maxwellian velocity distribution 
is assumed for the gas in each computational cell,
with the scale determined by the local gas temperature (see Mashchenko
\& Silich \cite{MS}).
No attempt was made to include the effects of turbulence. 
The model galaxy is assumed to have an inclination angle of $5^{\circ}$.
The position-velocity cuts along the minor axis of the projected shell have similar
appearance for such low inclination angles.

Figure~\ref{fig6} shows the model 
pV-diagrams of the shell produced by 500 consecutive SNe. The quantity
$S_{\rm HI}$ is plotted in all model pV-diagrams.
Two bright blobs elongated in the vertical direction are apparent in each pV-diagram. 
They represent the dense
walls of the shell expanding in the plane of the galaxy, i.e. perpendicular to
the line of sight at the adopted inclination angle of $5^\circ$.
If the shell has not yet broken out of the disk ($t=9$ and 18~Myr), 
these two blobs appear to close at their tops and form a complete 
elliptical ring. The relative amplitude
of $S_{\rm HI}$ around the ring is however varying by roughly an order
of magnitude, reflecting the difference in the HI column density along
the major axis of the projected shell --
there is much less gas on the approaching and receding sides of the shell
than on those expanding perpendicular to the line of sight.

The elliptical ring is virtually absent in the later expansion phase at $t=36$ and
45~Myr when the shell has already broken out of the disk (the lower frames
in Fig.~\ref{fig6}). 
Instead, the pV-diagrams are totally dominated by two high-intensity blobs. A 
few smaller patches seen between the blobs
represent the dense, cold clouds fragmented off the shell and pushed by
the dynamical pressure of the hot gas to a higher altitude of $\ge 1$~kpc.
The thickening of the shell after the blowout 
is also noticeable in the lower frames of Fig.~\ref{fig6}.
The same tendency is found in the pV-diagrams of the shell created by 1000
consecutive SNe. The complete ring is seen
only in the very early phase of the shell expansion at t=9~Myr and is virtually
absent after the blowout at $t=27$, 45, and 54~Myr.

\begin{figure}
    \resizebox{\hsize}{!}{\includegraphics{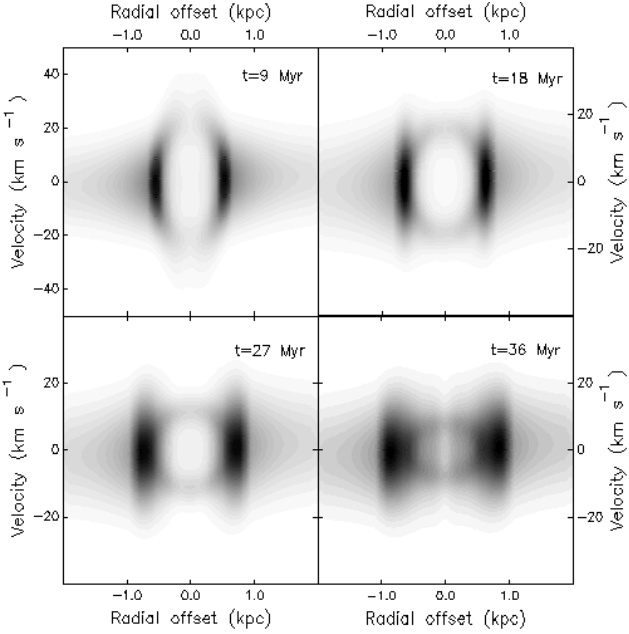}}
      \caption{The pV-diagrams of the shell produced by a single hypernova
      of the total energy of $5.0\times 10^{53}$~ergs. The pV-cuts are
      taken along the major axis of the model galaxy viewed at an inclination angle
      of $5^\circ$.}
         \label{fig7}
\end{figure}

\begin{figure}
    \resizebox{\hsize}{!}{\includegraphics{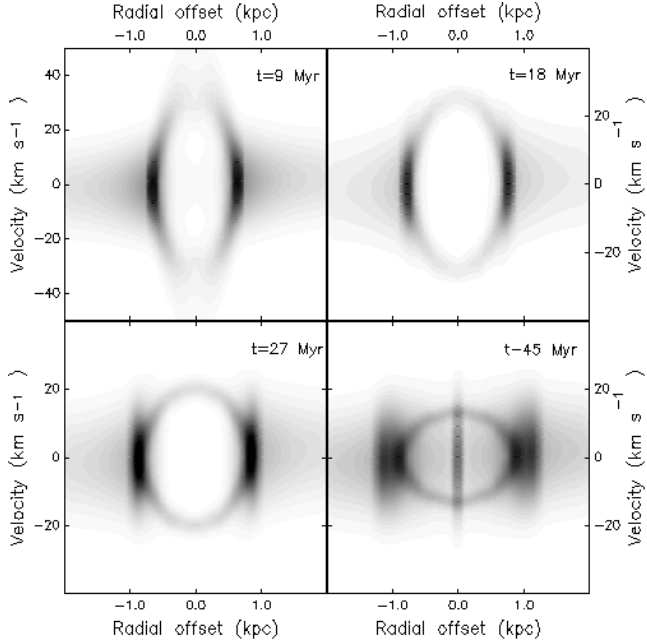}}
      \caption{The same as Fig.~\ref{fig7}, but for a single hypernova
      of the total energy of $10^{54}$~ergs.}
         \label{fig8}
\end{figure}

Figure~\ref{fig7} shows the pV-diagrams 
of the shell produced by a hypernova of total energy $5\times 10^{53}$~ergs.
The position-velocity cuts are taken along the major axis of the projected
shell at four different evolutionary times: $t=9$, 18, 27, and 36~Myr, i.e., before it
has expanded out to its maximum size ($t=9$ and 18~Myr) and after it has started to collapse
due to the radiative cooling of the shell interior. An inclination angle
$i=5^\circ$ is assumed. The elliptical hole surrounded by a ring of higher 
HI flux density $S_{\rm HI}$ is clearly visible in those pV-diagrams.
The HI flux $S_{\rm HI}$ from the ring is $\sim 7-10~ \mu Jy$, which
should, in principle, be detectable.
{\it Hypernova explosions tend to form complete shells that are clearly visible
in the pV-diagrams.}
This characteristic feature of a hypernova-driven shell 
is also present when one considers very energetic explosions.
For instance, Fig.~\ref{fig8}
shows the pV-diagrams of the shell produced by a hypernova of the total energy
of $10^{54}$~ergs. The elliptical hole surrounded by the ring is clearly seen
even when the shell has already started to collapse at $t=45$~Myr, as implied 
by the presence of a vertical bar at zero radial offset
(see the lower right panel in Fig.~\ref{fig8}).
Note the size of the shell in Figs.~\ref{fig7} and \ref{fig8}. 
It is twice the local Gaussian scale height $h$, which is $\sim 350$~pc at
a galactocentric radius of 0.5~kpc in our model galaxy. 

The comparison of Fig.~\ref{fig6} and Figs.~\ref{fig7},~\ref{fig8}  indicates that
the SNe-driven shell in its late expansion phase (when its radius
exceeds $1.5\,h$) may be distinguished from that produced by the single hypernova explosion
based on its appearance in the pV-diagrams.
The pV-diagrams of the hypernova-driven shell show the characteristic elliptical
ring in virtually any evolutionary phase of the shell, whereas
the pV-diagrams of the SN-driven shell do that only in the early expansion
phase, well before the blowout.
The present numerical simulations 
(see also Vorobyov \& Shchekinov 2004) show that the SN-driven shell breaks out of the disk
when its radius exceeds $\sim 1.5\,h$. 
As a consequence of the blowout, the SN-driven shell transforms into an
open cylinder and the pV-diagrams show two parallel blobs representing
the walls of the expanding cylindrical shell. 
On the contrary, the hypernova-driven shell appears to
survive the expansion even when its radius exceeds $2.0\,h$. As a result,
the characteristic elliptical ring representing the complete expanding
shell is clearly seen in the pV-diagrams.
Thus, if the pV-diagrams show an elliptical hole surrounded by a denser
ring, the radius of which is
greater than $1.5\,h$ (as in the lower frames of Fig~\ref{fig7} and
\ref{fig8}), this may indirectly point to a hypernova origin of the shell.

\subsection{Effects of inclination}
\begin{figure}
    \resizebox{\hsize}{!}{\includegraphics{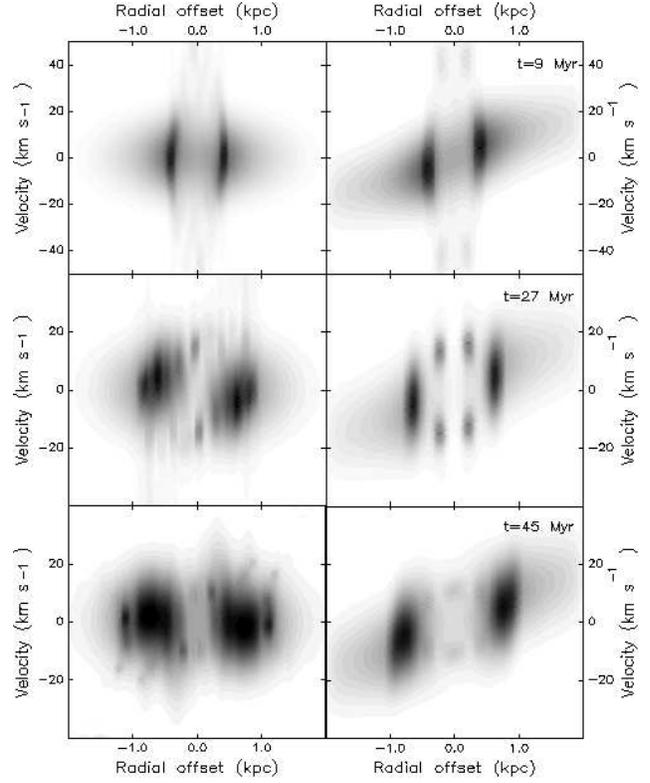}}
      \caption{The pV-diagrams of the shell created by 1000 consecutive
      SNe. The pV-cuts are taken along the minor (the left panels) and
      the major (the right panels) axis of the model galaxy viewed at an
      angle of $45^\circ$.}
         \label{fig9}
\end{figure}

The model pV-diagrams considered in Sect.~\ref{PVD} were constructed under
the assumption that the model galaxy is viewed nearly face-on, namely at an inclination
angle of $5^\circ$.
In this section we study how a higher inclination angle could influence the appearance
of the SN- and hypernova-driven shells in the pV-diagrams. Figure~\ref{fig9}
shows the model pV-diagrams of the shell created by 1000 consecutive SNe.
The pV-diagrams are constructed at three different evolutionary times $t=9$,
27, and 45~Myr after the beginning of the energy injection phase. The left
and right panels in Fig.~\ref{fig9} are the position-velocity cuts along the minor and
major axes of the projected shell, respectively. An inclination
angle of $i=45^\circ$ is adopted. The characteristic hole surrounded by a
ring of higher HI flux density (see Sect.~\ref{PVD})
is barely seen even in the very early expansion phase at $t=9$~Myr due
to the projection smearing at higher inclination angles.  
In the later expansion phase at $t=45$~Myr
the pV-cuts taken along the major axis show two elongated
blobs, the morphological centers of which are symmetrically displaced by $|\triangle v| \sim 7-8$~km~s$^{-1}$
with respect to $v=0$~km~s$^{-1}$. The displacement is
defined as $\triangle v=\pm v_{\rm rot}~\sin(i)$, where $v_{\rm rot}$ is the
rotation velocity of the SN-driven shell. The positive and negative $\triangle
v$ correspond to
the receding and approaching parts of the shell, respectively.
It is important to note here that similar pV-diagrams
are obtained for Ho~I  by Ott et al. (\cite{Ott}, see their Fig.~8).
A large displacement between the blobs with respect to the systemic 
velocity of $v_{\rm sys}=141.5$~km~s$^{-1}$
is clearly seen in the pV-cut taken along the major axis of Ho~I. 
However, we found it difficult to estimate the value of $\triangle v$ from the
observed pV-diagram shown in Fig.8 of Ott et al. (\cite{Ott}) due to its considerable
asymmetry (caused probably by an off-center location
of the stellar cluster responsible for the creation of the shell).
The modeled value of $\triangle v$ depends also on the adopted rotation curve
of Ho~I, which is poorly known (Ott et al. \cite{Ott}). 
Therefore, it is problematic to obtain the inclination angle of Ho~I
by comparing the modeled and observed $\triangle v$. 
Nevertheless, we conclude that Ho~I is viewed at an inclination
angle considerably higher than $i=5^\circ$, because the modeled pV-cuts of Fig.~\ref{fig6} obtained 
for an inclination angle of only $5^\circ$ do not show a noticeable
displacement.

\begin{figure}
    \resizebox{\hsize}{!}{\includegraphics{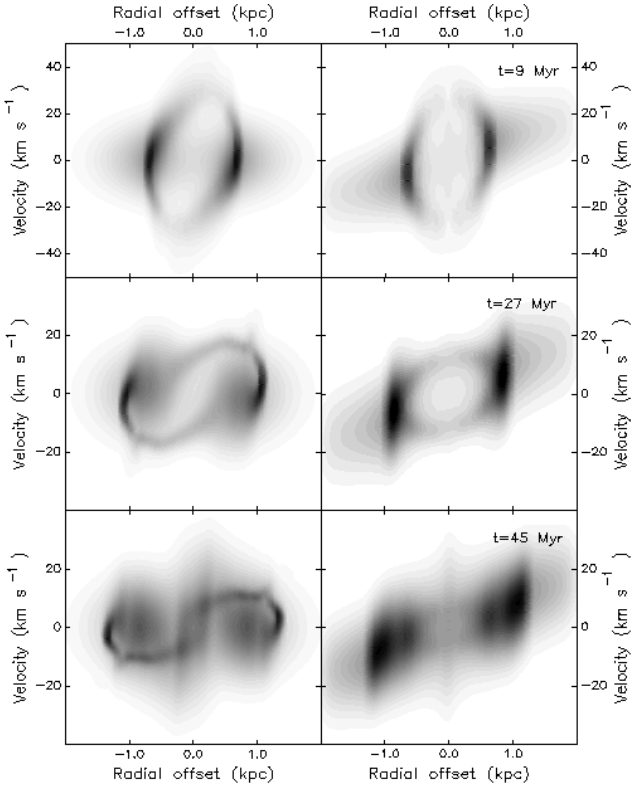}}
      \caption{The same as in Fig.\ref{fig9}, but for a single hypernova
      of the same total energy of $10^{54}$~ergs. The pV-cuts taken along
      the minor axis of the hypernova-driven shell show the characteristic
      S-shaped structure due to an expansion in the vertical direction.}
         \label{fig10}
\end{figure}

The pV-diagrams of the shell created by a single hypernova
explosion of the total energy $10^{54}$~ergs are shown in Fig.~\ref{fig10}.
The left panels give the pV-cuts along the minor axis of the projected
shell, whereas the right panels do those along the major axis.
The pV-diagrams of the shell created by a single hypernova explosion are
noticeably different from those created by multiple SNe. The hypernova-driven
shell does not break out of the disk even when its radius exceeds $2.0~h$. 
As a consequence, the shell preserves its elliptical form and 
the characteristic elliptical hole surrounded by the ring
of a higher HI flux intensity is apparent in the pV-cuts taken along the major
axis (the right panels in Fig.~\ref{fig10}) at $t=9$ and 27~Myr after the explosion.
The elliptical ring dissolves at $t\ge 45$~Myr due to the projection effect 
(note that in nearly face-on galaxies the elliptical
ring can still be resolved at that phase). In that later phase, the hypernova-driven shell can
barely be distinguished from that created by multiple SNe based on the
appearance in the pV-diagrams taken along the major axis of the projected
shell. Fortunately, the pV-cuts taken along the minor axis
of the hypernova-driven shell show the characteristic S-shaped structure.
The pV-cuts along the minor axis are sensitive only to the expansion/contraction
motion of the shell. As is seen in the lower panels of Fig.~\ref{fig3}, the hypernova-driven
shell  at $t=27-45$~Myr expands in the vertical direction, but shows the lack of expansion or contraction
motion in the galactic plane (except for the very late phase when the shell
has already started to collapse).  
As a result, the shell seen in the pV-diagrams as the elliptical ring 
transforms into the S-shaped structure in the later expansion phase.
Note that neither elliptical rings nor S-shaped structures are seen in
the observed pV-diagrams of Ho~I plotted in Fig.~8 of Ott et al. (\cite{Ott}).

\begin{figure}
    \resizebox{\hsize}{!}{\includegraphics{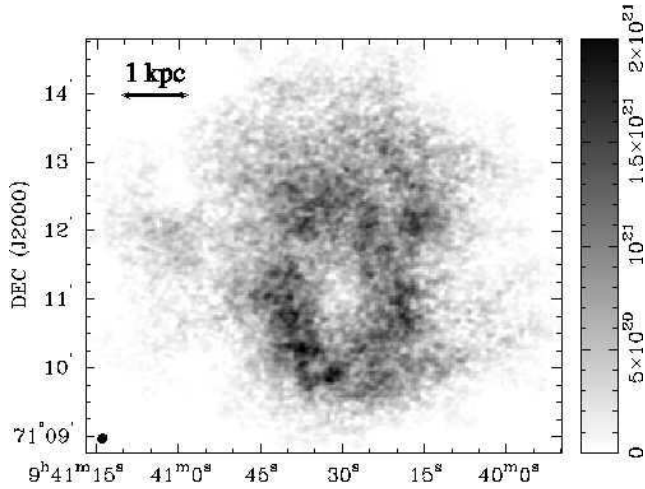}}
      \caption{Integrated HI emission of Ho~I. In the bottom left
      corner we show the half-power beam (8\farcs2 $\times$ 7.0\farcs).
      The grayscale represents the HI column density.
      The data have been obtained from the Very Large Array observations 
      (Ott et al. \cite{Ott}).}
         \label{fig11a}
\end{figure}

As was demonstrated in Fig.~\ref{fig5}, the HI rings created by 
multiple SNe have a much
higher contrast in the HI column
density  between the ring  and central depression,
${\cal K}=N_{\rm r}(HI)/N_{\rm c}(HI)$, than those created by a single hypernova
explosion of the same total energy.  
However, a higher inclination may considerably reduce the value of ${\cal
K}$ due to the projection smearing effect. For instance, the dotted-dashed line in Fig.~\ref{fig5} shows the
contrast ${\cal K}$ as a function of the ring radius $R$ produced by 1000 consecutive SNe, 
whereas the dotted
line  does that for the hypernova of the same total energy. An inclination
angle of $45^\circ$ is assumed. The ring-to-center contrast in $N(HI)$ produced by both multiple
SNe and  hypernova stays around ${\cal K}=2-3$. The only exception is
seen when the SN-driven shell expands out to $R \approx 0.7$~kpc and breaks
out of the disk. Then, the ring-to-center contrast reaches a maximum value of ${\cal K} \approx
8$. It is instructive to compare the ring-to-center contrast and size of HI rings produced
in our simulations (by both multiple SNe and single hypernovae)  with the observed
contrast ${\cal K}$ and size of the HI ring in Ho~I. Fig.~\ref{fig11a} shows
the observed HI map of Ho~I as obtained by Ott et al. (\cite{Ott}). The
diameter of the ring is $\sim 1.7$~kpc and  the ring-to-center contrast 
is ${\cal K}\sim 15-20$. Our simulations show (see Fig.~\ref{fig5}) that 
both multiple SNe and single hypernovae may form rings of up to 2~kpc in diameter.
However, single hypernovae fail to produce rings with the observed contrast
of ${\cal K}\sim 15-20$. This is true even in the most favorable case of near
zero inclination angle. 
Taking into account that both the observations
of Ott et al. and numerical modeling of Vorobyov et al. (\cite{Vor}) set
the lower limit on the inclination angle of Ho~I as $i\le 15^\circ$, the
formation of the HI ring with such a high contrast ${\cal K}\sim15-20$ by means of
a hypernova explosion seems highly problematic. Apparently, only
multiple SNe can produce HI rings with such a high ring-to-center contrast $\cal
K$.

\begin{figure}
    \resizebox{\hsize}{!}{\includegraphics{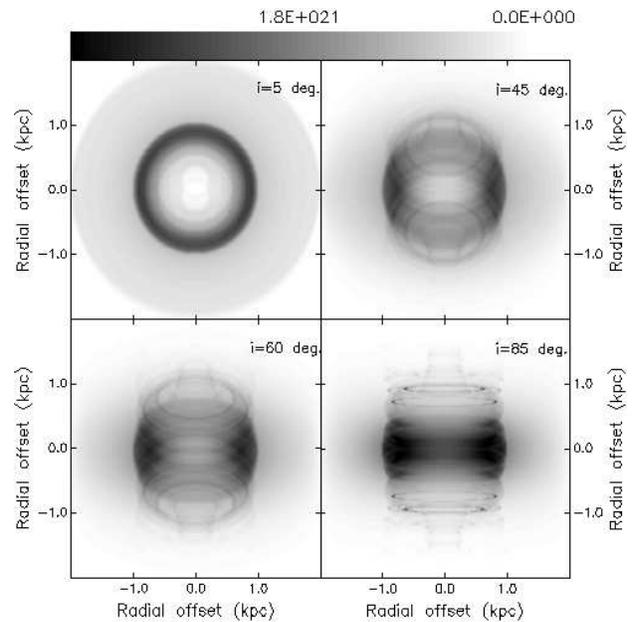}}
      \caption{The integrated HI image of the ring produced by 500 consecutive
      SNe. The inclination angle is indicated in the right upper corner.
      The scale bar is in cm$^{-2}$. {The origin of narrow rings is an assumption
      of axisymmetry in our numerical simulations.}} 
         \label{fig12}
\end{figure}

The integrated HI image  of the shell created by multiple 
SNe  appears in nearly face-on galaxies as 
a ring of elevated HI density surrounding the central depression.
For instance, the left upper panel in Fig.~\ref{fig12} shows the integrated HI image
of the shell created by 500 consecutive SNe. The HI image is taken at $t=45$~Myr after the beginning
of the energy input phase, when the shell has already broken out of the
disk. An inclination angle of $i=5^\circ$ is assumed. As is obviously seen, 
the blown-out shell in nearly face-on galaxies appears
as a ring with a high ring-to-center HI column density contrast ($\sim 25-30$).
There is little azimuthal variation in the HI column density around the
ring. 
However, the integrated HI image of the blownout shell observed in substantially 
inclined galaxies shows a considerable variation in the HI column density around the
ring. For instance, at 
higher inclination angles of $45^\circ$ and $60^\circ$ the relative amplitude
of azimuthal variations
in $N(\rm {HI})$ around the ring is $\approx 2$ and $\approx 3$, respectively.
In this case, the integrated HI image is dominated by two kidney-shaped
density enhancements and shows a mild central minimum.
It is interesting to note that the Sextans~A and M81~dwA
dwarf irregular galaxies have a somewhat similar HI distribution (see Skillman et
al. \cite{Skillman}, Stewart \cite{Stewart}). In
the case of Sextans~A, the relative amplitude of azimuthal variations in
the HI column density around the ring  
is $\approx 2.5-3$ (according to Skillman et al. \cite{Skillman}) and the inclination angle
determined from the outer HI contours is $i=36^\circ \pm 4^\circ$ (assuming
the circular symmetry).  
Note that the narrow ring-shaped structures seen in Fig.~\ref{fig12} are in fact the dense fragments
of a broken-up shell, which appear as rings rather than dense clumps due
to the axisymmetric nature of our numerical simulations.  
In edge-on galaxies, the integrated HI image would resemble a `dumbbell',
as shown in the right lower panel of Fig.~\ref{fig12}.
The absence of noticeable azimuthal variations in HI column density
around the HI ring in Fig.~\ref{fig11a} suggests that
Ho~I is viewed at an inclination angle considerably smaller than $i=45^\circ$.
This is in agreement with previous estimates by Ott
et al. (\cite{Ott}) and Vorobyov et al. (\cite{Vor}), who found that Ho~I
has an inclination angle of $\la 20^{\circ}$. 

\subsection{Off-plane explosions}
\label{offplane}
A general assumption made in the above simulations of both SN-driven
and hypernova-driven shells was that the energy is released exactly in
the midplane of the model gas disk. However, taking into account a considerable
thickness of the gas disks in dIrr's, the occurrence of a hypernova or massive
stellar cluster at a position moderately offset with respect to 
the galactic midplane, appears plausible. In this section we investigate how
the off-plane explosions may influence the appearance of the shells in
the pV-diagrams. 

\begin{figure}
    \resizebox{\hsize}{!}{\includegraphics{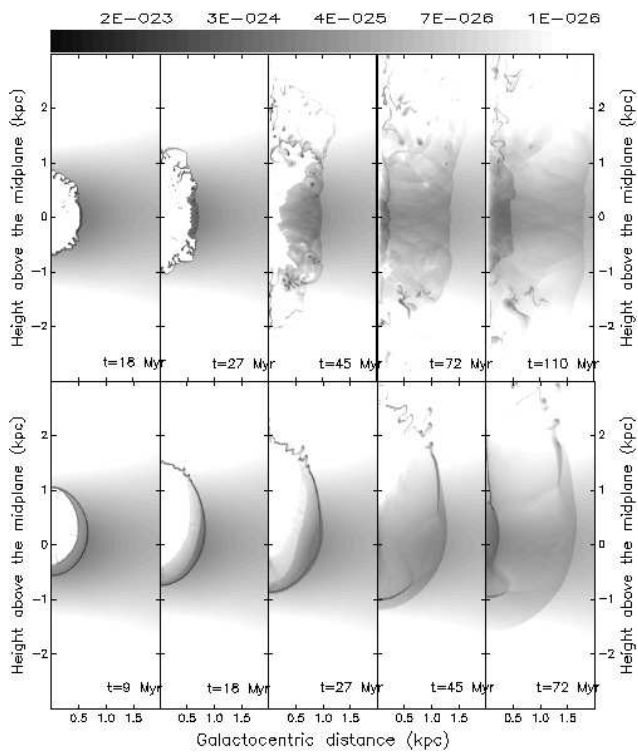}}
      \caption{Off-plane explosions. Temporal evolution of the gas volume density distribution
      after the release of $1.0\times 10^{54}$~ergs of thermal energy at $r=0,z=100$~pc. 
      The upper panels correspond to 1000 consecutive SN explosions over
      30~Myr, while the lower panels show the impact of a single hypernova
      of the same total energy of $10^{54}$~ergs. The scale bar is in gm~cm$^{-3}$.}
         \label{fig13}
\end{figure}

If the stellar cluster is located at a moderate height of $z=100$~pc ($\sim
1/3~h$) above the midplane of the galaxy, the temporal evolution of the gas volume density
is found to be similar to that obtained 
in Sect.~\ref{SNGRB} for the midplane stellar cluster.
For instance, the upper panels in Fig.~\ref{fig13} 
show five temporal snapshots
of the gas volume density distribution created by the energy release of
1000 consecutive SNe. A more intensive blowout takes place in the upper
part of the galaxy where the stellar cluster is nested. 
However, the pV-diagrams and ring-to-center contrast in HI column density
$\cal K$ are only slightly affected by this mild asymmetry.

The temporal evolution of the shell created by a single hypernova
that is moderately offset with respect to the midplane of the galaxy produces
some noticeable differences. The lower panels in Fig.~\ref{fig13} 
give five temporal snapshots of the gas volume density distribution created
by the hypernova of total energy $10^{54}$~ergs that is placed 100~pc above
the midplane of the galactic gas disk. The egg-like shell is formed after
$\approx 15$~Myr, the upper part of which retains a relatively smooth appearance
for several tens of Myr before it breaks due to a Rayleigh-Taylor instability
and dissolves at $\approx 70-80$~Myr after the explosion. 
Note that in the case of multiple SNe such smooth kpc-sized
arcs of dense material are never formed in our numerical simulations. Instead,
inspection of Figs.~\ref{fig1}-\ref{fig3}, and Fig.~\ref{fig13} indicates
that multiple SNe tend to produce patchy filaments and dense clumps of
a few hundred parsecs in size (except for the very late phase when the
shell has already collapsed). Indeed, a Rayleigh-Taylor instability, the
effect of which is strengthened
by the preceding Kelvin-Helmholtz instability, efficiently fragments the
SN-driven shell when the latter breaks out of the gas disk. Longer fragmentation
times of hypernova-driven shells as compared to fragmentation times of
shells created by multiple SNe were also found in the model of Efremov et al. (\cite{EEP}).
It is interesting to mention in this context the W4 HII
region in the Perseus arm of the outer Galaxy. Dennison et al. (\cite{Dennison}) 
have observed a highly elongated shell of $\rm H\alpha$ emission, the base
of which is located near the W4 HII region. The $\rm H\alpha$ shell appears
to close far above the Galactic plane and is interpreted by Basu et al.
(\cite{Basu})
to be the dense ionized wall of swept-up gas surrounding a bubble of hot
rarefied gas created by the stellar winds from the massive stars.
However, the apparent smooth appearance of the $\rm H\alpha$ shell implies
that the Rayleigh-Taylor instability is suppressed there, possibly due
to the stabilizing effect of a swept-up tangential magnetic field in the
shell (Komljenovic, Basu, \& Johnstone \cite{KBJ}).
An alternative explanation may involve a shell formed by a hypernova explosion.
In that case, the characteristic growth time of a Rayleigh-Taylor instability in the expanding shell 
is much longer than in the case of multiple SNe, and the shell may survive
for a longer time and expand to a higher altitude.

\begin{figure}
    \resizebox{\hsize}{!}{\includegraphics{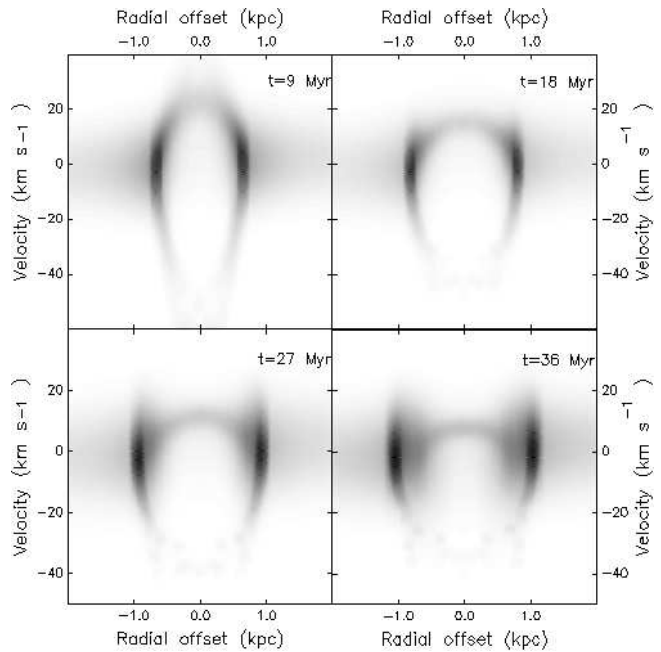}}
      \caption{Off-plane explosion. The pV-diagrams of the shell created by a single hypernova
      of the total energy of $10^{54}$~ergs. The pV-cuts are taken along the major 
      axis of the model galaxy viewed at an
     inclination  angle of $5^\circ$.}
         \label{fig14}
\end{figure}

Off-plane hypernova explosions may leave noticeable signatures in the pV-diagrams.
While the upper part of the egg-like shell is thin and contains a small fraction
of the total gas mass, the lower part of the shell accumulates a larger amount
of gas mass and becomes clearly visible in the pV-diagrams. Figure~\ref{fig14}
shows the pV-cuts taken along the major axis of the 
shell created by a single hypernova of the total energy of $10^{54}$~ergs,
which is placed 100~pc above the midplane of the galaxy.
An inclination angle of 
$5^\circ$ is adopted. The bottom of the egg-like shell
appears in the pV-diagrams as a bright arc connecting two vertical blobs.
The latter are the walls of the shell expanding in the horizontal direction.
The top of the shell is clearly visible only in the very early expansion
phase at $t\le 10$~Myr. However, it could be seen in $H\alpha$ emission,
if a powerful ionizing source is located inside the shell, as it might be
the case for W4 HII region discussed above.

\section{High velocity clouds}
\label{cloud}
In this section we numerically study the vertical impact of high velocity clouds 
as a potential mechanism of the giant HI ring formation in Ho~I.
The numerical simulations of multiple SNe in Sect.~\ref{SNGRB} 
have shown that it requires
$>2.0 \times 10^{53}$~ergs of released energy to form HI rings with 
HI ring-to-center contrast 
${\cal K}>10 $. Hence, we consider only very energetic HVCs that could
potentially deposit a comparable amount of kinetic energy to the interstellar
medium.  We have performed a parameter study assuming that HVCs cover a range of velocities 
$200~{\rm km~s}^{-1}\le v_{\rm \scriptstyle HVC}\le 300~{\rm km~s}^{-1}$
and masses $2.5 \times 10^{5}~M_\odot \le M_{\rm \scriptstyle HVC} \le 2.5\times
10^6~M_\odot$. Thus, the kinetic energies of HVCs range from $1.0 \times
10^{53}$ to $2.5 \times 10^{54}$~ergs. Clouds are assumed to have a constant density
and be either spherical or cylindrical in shape. The initial position of
the cloud center is 3.2~kpc above the midplane.

\begin{figure}
    \resizebox{\hsize}{!}{\includegraphics{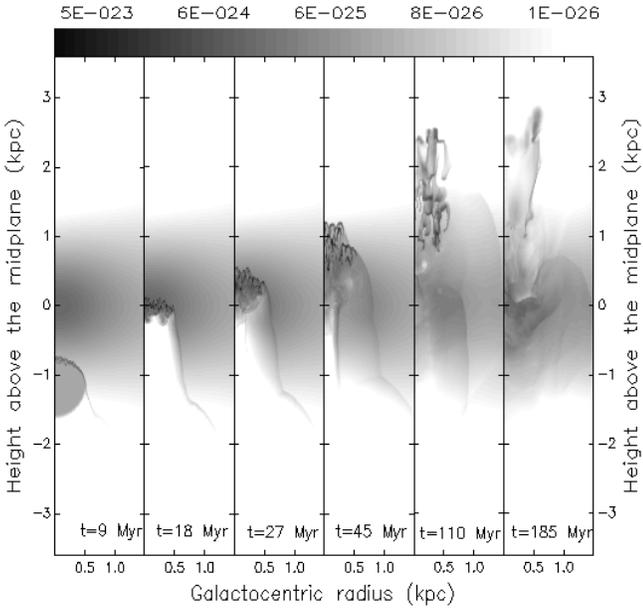}}
      \caption{The distribution of the gas volume density created by the
      collision of a cloud of $M_{\rm \scriptstyle HVC}=2.5 \times 10^{6}~M_\odot$ 
      and $v_{\rm \scriptstyle HVC}=200$~km~s$^{-1}$ with the galactic
      gas disk.}
         \label{fig15}
\end{figure}

Figure~\ref{fig15} shows the temporal evolution of the distribution of the
gas volume density produced by the impact of a cloud of $M_{\rm \scriptstyle
HVC} = 2.5 \times 10^6~M_\odot$ and  $v_{\rm \scriptstyle HVC}=200$~km~s$^{-1}$.
The total kinetic energy of the cloud is thus equal to 
$1.0\times 10^{54}$~ergs. The maximum column density $N({\rm
HI})$ of the cloud measured along its diameter 
is $4.3~\times 10^{20}$~cm$^{-2}$. The impinging HVC creates a hot
rarified bubble by pushing matter toward the galactic plane.
The bubble does not form a complete sphere and fills out 
$\sim 40$~Myr after the onset of the collision. The HVC continues
to push the gas along its way and eventually breaks through the disk. At
this stage ($t\approx 110$~Myr), the initial
shape of the HVC is completely lost due to a combined action of Rayleigh-Taylor
and thermal instabilities.
The HVC emerges above the galactic plane as a complicated network of dense filaments, 
clumps, and  arcs.
The formation of dense filaments and clumps was also reported in numerical
simulations by Franco et al. (\cite{Franco}).
It takes another 70~Myr for this remnant to fall back onto the disk and
create another bubble, though less prominent than the first one.

\begin{figure}
    \resizebox{\hsize}{!}{\includegraphics{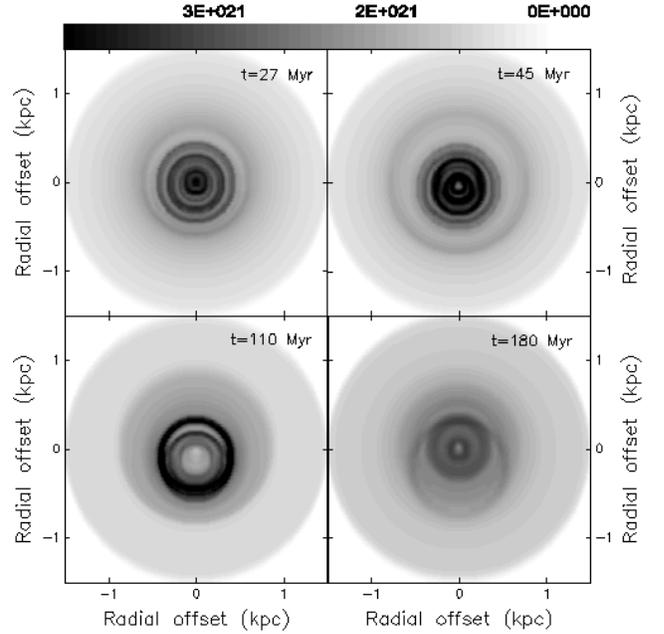}}
      \caption{The integrated HI image of the gas volume density distribution
      shown in Fig.~\ref{fig15}. An inclination angle of $5^\circ$ is adopted.
      The scale bar is in cm$^{-2}$.}
         \label{fig16}
\end{figure}

Although the kinetic energy of the impinging cloud is high ($10^{54}$~ergs)
and the cloud produces a considerable disturbance in the gas disk of the
target galaxy, we have found that it is rather ineffective in creating giant HI
rings similar to that observed in Ho~I.
The HVC pushes the matter mostly in the vertical direction so that the HI image of a
nearly face-on galaxy does not show a considerable radial variation
in the HI column density within the impact region. Indeed, Fig.~\ref{fig16}
shows the integrated HI image of the gas volume density distribution shown
in Fig.~\ref{fig15}. An inclination angle of $5^\circ$ is assumed.
The formation of a ring-like structure is only seen in the late collision
phase ($t\sim 110$~Myr), when the cloud has plunged through the gas disk
and pushed a considerable fraction of the disk gas to an altitude $z\ga 1$~kpc. 
Most of the gas constituting this remnant retains 
the orbital momentum it had before the collision.
Since the horizontal gravity is much less at $z\ga 1$~kpc than in
the plane of the galaxy, the lifted gas starts spiraling away on ballistic
orbits and it can be observed in nearly face-on galaxies as a dense HI ring surrounding the central
depression (see the left lower panel in Fig~\ref{fig16}). The maximum size of such an HI ring (diameter
$\sim 0.8$~kpc) is at least twice smaller than that found in our numerical simulations of multiple SNe
and hypernova explosions of the same total energy. Note that the
diameter of the HI ring in Ho~I is $\sim 1.7$~kpc. We find that most of the HI rings formed
by the impact of HVCs have diameters less than $0.5$~kpc.  The HVCs of a smaller
size and higher HI column density, but the same kinetic energy, 
tend to form smaller rings. The size of the HI ring depends weakly on the shape of the cloud.
HVCs with the total kinetic energy below $2\times 10^{53}$ do not form
noticeable HI rings. HI rings with a diameter
of $\ge 0.8$~kpc are never formed in our simulations for the considered
upper limit of HVC kinetic energy, $2.5\times 10^{54}$~ergs.

A more dramatic difference between the HI rings formed by the vertical
impact of
HVCs and those created by multiple SNe is seen in the values of $\cal K$. 
As is obviously seen from Fig.~\ref{fig5},
1000 consecutive SNe can create rings with $\cal K$ as high as 100, whereas
Fig.~\ref{fig16} indicates that the impact of a cloud of the same kinetic 
energy generates the ring with ${\cal K}\approx 4.5$.
Note that $\cal K$ in Holmberg~I is $\sim
15-20$ (Ott et al. \cite{Ott}).
The same tendency is found for other values of released energy: multiple
SNe produce rings with roughly 20-30 times
higher values of $\cal K$ than 
does the impact of a cloud of the same total energy. On the other
hand, hypernova explosions form rings with $\cal K$ roughly a factor of 2 higher
than those created by the clouds of the same kinetic energy.
The difference in $\cal K$ between the rings formed by the three mechanisms
considered above is smeared out as one considers higher inclination angles
of $i>45^\circ$.

\begin{figure}
    \resizebox{\hsize}{!}{\includegraphics{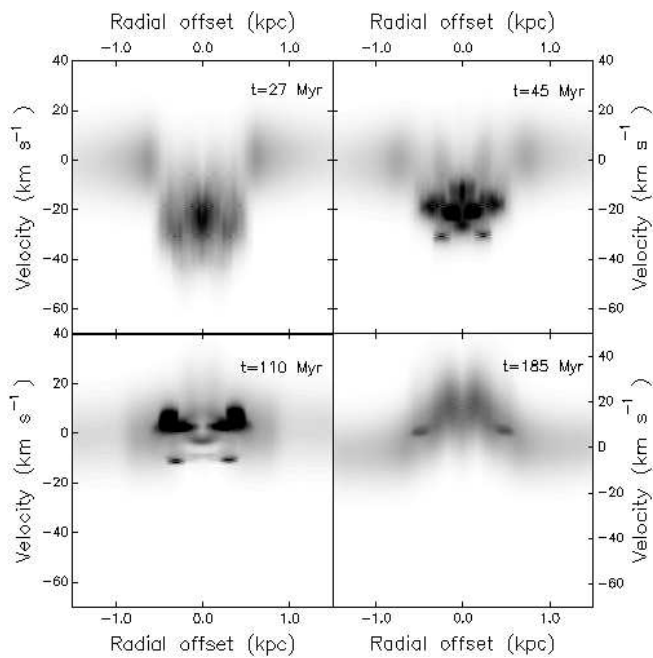}}
      \caption{The pV-diagrams of the gas volume density distribution shown
      in Fig.~\ref{fig15}. The pV-cuts are taken along the major axis of
      the projected galaxy. An inclination angle of $5^\circ$ is assumed.}
         \label{fig18}
\end{figure}

The impact of HVCs should leave prominent features in the pV-diagrams,
which could be used to distinguish between the rings formed by the impact
of HVCs and those formed by multiple SN and hypernova explosions. Indeed,
Fig.~\ref{fig18} shows the pV-cuts taken along the major axis of the
galaxy experiencing the collision with a cloud of $M_{\rm \scriptstyle
HVC} = 2.5 \times 10^6~M_\odot$ and  $v_{\rm \scriptstyle HVC}=200$~km~s$^{-1}$
An inclination angle of $5^\circ$ is adopted.
The open bubble produced by the impinging cloud is clearly seen in the
upper left panel of Fig.~\ref{fig18}. The bubble starts filling out after $t=45$~Myr.
However, the cloud is still clearly visible in the pV-diagrams at that
time as a dense irregular clump moving towards the observer. The velocity of the cloud
noticeably decreases during the passage from $v_{\rm \scriptstyle HVC}\approx 200$~km~s$^{-1}$ at
the onset of collision to $v_{\rm \scriptstyle HVC}\approx 20$~km~s$^{-1}$
at the time when the cloud emerges on the other side of the gas disk.
The pV-diagram at this time ($t=110$~Myr) is dominated by two bright and compact clumps,
which represent the cold dense gas compressed and pushed by the cloud to
an altitude of $z\approx 1-2$~kpc. The depression
in the HI flux seen in the pV-diagram near the galactic center is due to the HI
ring formation discussed above.  Note that these two clumps are noticeably different in
shape from those observed in the pV-diagrams of the shells formed by multiple
SN and hypernova explosions (see Figs.~\ref{fig6}-\ref{fig8}).
The second bubble formed by the remnant falling back onto the disk is clearly
seen in the lower right panel of Fig.~\ref{fig18}.

It is obviously seen that neither the pV-diagrams nor the size and/or ring-to-center
contrast $\cal K$ obtained in the simulations of the vertical collision of HVCs with
the galactic disk can account for what is actually
observed in Ho~I. We thus conclude that multiple SN explosions 
remain the most plausible explanation for the  origin of the giant
HI ring morphology in Ho~I.

\section{Discussion}
If SNe are the origin of HI rings surrounding HI holes, the presence 
of a remnant stellar cluster is expected. Such stellar clusters are not always observed within HI
rings or holes. For instance, Rhode et al. (\cite{Rhode}) found that
several HI holes in the dwarf irregular galaxy Holmberg~II lack any stellar
cluster counterparts. However, in the case of Holmberg~I, the $B$-band
magnitude as well as the $U-B$ color of the optical emission from the giant HI
hole indicates that a stellar cluster of $\sim 2\times 10^5~M_{\odot}$
may be present there (Vorobyov et al. \cite{Vor}). 

The high density clumps produced in our simulations due to the fragmentation of
SNe-driven shells are usually not self-gravitating. The masses of these clumps
are slightly below the Jeans mass $M_{\rm J}=5\, c_{\rm s}^2 d/(4G)$, 
where $d$ is the size of the clump. This expression is derived 
from the virial theorem assuming  spherical clumps. This is consistent with
a low $H\alpha$ luminosity of Ho~I: $L_{\rm H\alpha}=4.3 \times 10^{38}$~ergs
s$^{-1}$ (Miller \& Hodge \cite{Miller}). This also implies that
SNe alone have difficulty in producing self-gravitating clumps in dwarf
irregular galaxies, and probably
an external agent such as ram pressure or the collision of two or more SNe-driven
shells is needed to induce star formation. In massive disk galaxies, radiative
cooling is usually more efficient than in dwarf irregulars due to a 
higher metallicity. Therefore, dense
clumps are more likely to become Jeans unstable.
On the other hand, the most energetic HVC's with kinetic energy $\ga 1.0 \times 10^{54}$~ergs
can indeed produce dense clumps that are Jeans unstable. However, such energies
are rather exceptional and hence this mechanism cannot account for most star formation
in dwarf irregulars.

There are a few assumptions inherent to the model that need further discussion.\\
{\it  The interstellar medium distribution in Ho~I.}  The solution of the
steady state momentum equation in Sect.~\ref{model} produces a smooth initial gas distribution,
which may be only a first order approximation to the more realistic inhomogeneous
interstellar medium of dwarf irregulars. The influence of the deviations
from a smooth ISM distribution in the gas disk on the formation and evolution
of  HI shells deserves a separate detailed study. We expect here
that the effect of small scale inhomogeneities   with sizes much smaller than
that of the HI shell should be cancelled out as the shell forms. Conversely,
large scale inhomogeneities with sizes comparable to that of the HI shell
could  influence its shape and evolution, making it easier/harder to break
out of the disk. In particular, large scale inhomogeneities may reduce the
smooth appearance of hypernova-driven shells discussed in Sect~\ref{offplane}.
However, considering the size of the HI ring in Ho~I ($\sim 1.7$~kpc), we do not expect
inhomogeneities of a such scale to be present in the initial gas distribution
of this galaxy. \\
{\it The vertical scale height of the stellar disk and gas velocity dispersion.} 
The actual value of the vertical scale height $z_{\rm s}$ of the
stellar disk is difficult to estimate, since Ho~I has a low inclination
angle. We have adopted a value of $z_{\rm s}=300$~pc, which is typical
for dwarf irregular galaxies. A moderate variation $\pm 100$~pc in the value of $z_{\rm s}$ 
has a minor influence on the gas distribution and, hence, on our main results.
The HI velocity dispersion of Ho~I is measured by Ott et al. (\cite{Ott},
see their Fig.~10).
A moderate value $\sigma_{\rm HI}\approx 9$~km~s$^{-1}$ is found for most of Ho~I
disk, which is typical for gas-rich, quiet dwarf galaxies (Stil \cite{Stil}) 
and is used in our numerical simulations.
A small portion of gas disk in the northwestern part of Ho~I has, 
however, a slightly higher value of $\sigma_{\rm HI}\approx12$~km~s$^{-1}$. 
Such a variation in $\sigma_{\rm HI}$ is not expected to influence our main results. 
On the other hand, the measured $\sigma_{\rm HI}$ 
represents the cold component of the gas
disk only. If a massive hot gas component is present in Ho~I, the 
velocity dispersion $\sigma_{\rm HI+HII}$ of the mixture of hot 
and cold gas components may be
significantly greater. An increased gas velocity dispersion makes the gas disk
thicker. For instance, for an adopted value of $\sigma_{\rm HI+HII}=20$~km~s$^{-1}$,
the Gaussian scale height of the model gas disk $h$  becomes roughly three times 
larger than that shown by the dotted line in Fig.~\ref{fig0}. 
However, $h\ga 1$~kpc is not expected in Ho~I, 
because a SN-driven shell with radius $r\approx 0.7-0.8$~kpc would not 
blow out of the disk. 
Therefore, we would not observe such a high column density contrast 
(${\cal K}\approx 15-20$) between the HI ring and the central depression.
Furthermore, Ho~I is a quiet dwarf galaxy 
with an average star formation rate of only $0.004~M_{\odot}$~yr$^{-1}$ (Miller
\& Hodge \cite{Miller})
and we do not expect a large amount of hot gas
to be present there. Hence, we conclude that the integrated 
gas velocity dispersion of Ho~I is not expected to deviate much from the
adopted value of $\sigma=9$~km~s$^{-1}$.

{\it Internal structure and geometry of HVCs.} The shape of HVCs (i.e.
spherical or cylindrical) is found
to have a very minor influence on our results, which is agreement with
many previous studies of HVC collisions with  a gas disk (e.g. Comeron \&
Torra \cite{Comeron}). The assumption of an initially homogeneous cloud is
certainly an idealized case, since the observations show that 
HVCs are very clumpy and hierarchically structured (e.g. Wakker \& Schwarz
\cite{Wakker}). It may be that the remnant of the initially homogeneous cloud
that emerges at the other side of the galactic plane at $t\approx 110$~Myr in
Fig.~\ref{fig14} is a better example of an HVC prior to the collision.
Inhomogeneous structure further reduces the ability of an HVC to produce
large HI rings.

\section{Conclusions}
\label{sum}

We have numerically studied the formation of a giant HI ring as is observed
in the Holmberg~I dwarf irregular galaxy (Ott et al. \cite{Ott}).
The following three energetic mechanisms with the total released energy
of $\la 10^{54}$~ergs have been considered: multiple SNe, a 
hypernova explosion associated with a gamma ray burst, and the vertical
impact of a HVC. 
The shells created by those mechanisms are seen in the integrated HI image
of a nearly face-on galaxy as the HI rings of various sizes and ring-to-center contrasts
$\cal K$ in the HI column density.
We find the following differences among the HI rings formed by those mechanisms:
\\
$\bullet$ Multiple SNe form rings with the highest ring-to-center contrast
${\cal K}\la 100$.  A single hypernova
and an HVC form the rings with a much smaller values 
of ${\cal K}\la 10$ and ${\cal K} \la 5$, respectively.
This is due to the difference in the dynamics of the shells produced by
multiple SNe and those created by a hypernova: multiple SNe are
much more effective in producing blowout due to a combined action of
Kelvin-Helmholtz and Rayleigh-Taylor instabilities
than are single hypernovae of the same total energy. 
The vertical impact of an HVC is found to be ineffective in evacuating the gas in the
radial direction and hence in creating HI rings with a high ring-to-center
contrast. It is important to note that the difference in $\cal K$ found among 
three energetic mechanisms 
smears out at an inclination angle of $i\ge
45^\circ$ due to the projection effect.\\
$\bullet$ Both multiple SNe and a hypernova can account for the
observed size of the HI ring in Ho~I (diameter $\sim 1.7$~kpc). 
The maximum size of the ring that an HVC is found
to form in our numerical simulations is $\sim 0.8$~kpc.
We thus conclude that {\it only multiple SNe can reproduce both
the size {\rm (diameter $\sim 1.7$~kpc)} and the ring-to-center contrast in
the HI column density {\rm (${\cal K}\sim 15-20$)} of the HI ring in Ho~I.}

We construct the model position-velocity diagrams and find that they can be used to distinguish
between HI rings produced by different energetic mechanisms.
The model pV-diagrams of the blownout shell produced by multiple SNe 
are similar to those observed in Ho~I and show two symmetrically displaced,
elongated blobs representing the walls of the shell.
On the other hand, the model pV-diagrams of
the shell created by a hypernova explosion show a characteristic
elliptical ring-like structure, which is not seen in the observed pV-diagrams of Ho~I.
The impact of HVCs leaves prominent features in the model pV-diagrams that
are also not found in the observed pV-diagrams of Ho~I.
Those features include dense irregular clumps  and a void representing the impinging
cloud and the bubble formed by the impact.
We thus conclude that {\it multiple SNe 
appear to be the most plausible energetic mechanism
of the HI ring formation in Ho~I.}

Our numerical simulations indicate that the appearance of the SN-driven
shell in dwarf irregulars may depend on the inclination of the galaxy.
In nearly face-on galaxies the blownout shell appears in the integrated
HI image as an HI ring with a deep central minimum. There is little azimuthal variation
in the HI column density around the ring. 
The integrated HI image of the blownout shell in considerably inclined galaxies 
($i\ge 45^\circ$) is instead dominated by two
kidney-shaped enhancements and shows a mild central minimum, which is similar to what
is found in Sectans~A and M81~dwA (Skillman et
al. \cite{Skillman}, Stewart \cite{Stewart}). 
In nearly edge-on galaxies the integrated HI image of the blownout shell
resembles a `dumbbell'.

The offplane explosion of a hypernova creates the egg-like shell
that can preserve its shape for a considerably longer time than the
shell created by multiple SNe of the same total energy.
This may explain the smooth appearance of the highly elongated $H\alpha$ shell
observed by Dennison et al. (\cite{Dennison}) in the W4~HII region of the Perseus arm.

\section*{ Acknowledgments}
The authors are thankful to the referee, Prof. J. Palou{\v s}, 
for valuable suggestions and comments that helped improve the manuscript.
This work was supported by the NATO Science Fellowship Program administered
by the Natural Sciences and Engineering Research Council (NSERC) of Canada.
EV is grateful to the staff of the Department of Physics and Astronomy,
University of Western Ontario for their hospitality.

\end{document}